\documentclass[prl,twocolumn,amsmath,amsfonts,amssymb,showpacs,final]{revtex4}
\usepackage{graphicx,graphicx,epsfig}

\newif\ifpdf 
\ifx\pdfoutput\undefined
\pdffalse 
\else
\pdfoutput=1 
\pdftrue
\fi

\usepackage{graphicx}
\usepackage{bm}

\newcommand{\la}{\langle}
\newcommand{\ra}{\rangle}

\begin{document}
\title{The unified geometric theory of mesoscopic stochastic pumps and
  reversible ratchets}

\author{N.A. Sinitsyn}
\email{nsinitsyn@lanl.gov}
\author{Ilya Nemenman}
\email{nemenman@lanl.gov}
\affiliation{Center for Nonlinear Studies and Computer, Computational
  and Statistical Sciences Division, Los Alamos National Laboratory,
  Los Alamos, NM 87545 USA}

\pacs{03.65.Vf,05.10.Gg,05.40.Ca}

\begin{abstract}
  We construct a unifying theory of geometric effects in mesoscopic
  stochastic kinetics. We demonstrate that the adiabatic pump and the
  reversible ratchet effects, as well as similar novel phenomena in
  other domains, such as in epidemiology, all follow from geometric
  phase contributions to the effective action in the stochastic path
  integral representation of the moment generating function of
  particle fluxes. The theory provides a universal technique for
  identification, prediction and calculation of pump-like phenomena in
  an arbitrary mesoscopic stochastic framework.
\end{abstract}

\date{\today}

\maketitle

{\em Introduction}. A number of effects in classical nonstationary
statistical physics, such as the reversible ratchet
\cite{astumian-02pt, parrondo-98,shi} and the adiabatic pump effects
\cite{westerhoff-86, sinitsyn-07}, are known (or anticipated) to have
a geometric origin.  Examples and applications can be found in a
variety of fields, including electronic metrology \cite{metrology},
cell motility \cite{wilczek-88}, ion pumping through a cell membrane
\cite{tsong}, and dissipative chemical kinetics \cite{kagan-91}.  The
distinct feature of these effects is that, under a slow periodic
perturbation, transport coefficients are not a simple average of those
in the strict static approximation, but contain an extra component,
which changes its sign under a time-reversal of the perturbation.
Although these effects have been well-studied in a variety of fields
(see \cite{reimann-02} for reviews), a general theory, which clearly
disambiguates the pump (or ratchet) currents from other nonequilibrium
transport, provides a unified view of disparate pump-like phenomena,
and suggests universal quantitative methods for calculation of moments
of pump fluxes, is still missing.

We address this problem using the recently introduced stochastic path
integral representation of the moment generating function (mgf) of
fluxes in mesoscopic stochastic systems \cite{pilgram-03,pilgram-04}.
In this letter, we demonstrate that the stochastic path integral
technique can be employed to calculate moments of pump fluxes in a
general stochastic driven system in the mesoscopic (many particles)
and the adiabatic (slowly varying external parameters) regimes, and
that it makes a clear distinction between the pump fluxes and other
currents by relating the former to a geometric phase contribution to
the flux mgf. Our theory clarifies the connection between the
reversible ratchet and the adiabatic pump and allows to identify
similar effects in new contexts, which we demonstrate in a specific
scenario from epidemiology.

{\em Pump current from particle exclusion.} 
Let two absorbing states $S$ and $P$ (substrate and product in a
Michaelis-Menten enzymatic reaction \cite{MM,sinitsyn-07}, distinct
cellular compartments, neighborhoods of a city), exchange particles
(molecules, humans) via an intermediate system $B$ (bin, enzyme,
membrane channel, transportation hub). 
Our goal is to find the $S\to P$ flux $J$ and its fluctuations on time
scales much larger than the fluctuation time in the bin $B$, assuming
the mesoscopic regime with a large typical number of particles in the
bin, $N\gg1$.

Particles interact, and the in- and out-going transition rates can
depend on the number of particles in the bin, $N(t)$. The simplest
example of this kind is when the bin has a finite size, so that $N\le
N_B={\rm const}<\infty$. Then the in-rates are proportional to the
number of empty spaces in the bin, while the per-particle out-rates
are not affected by the occupancy.  The full kinetic scheme is
\begin{enumerate}\itemsep 0mm \partopsep0mm\topsep 0mm \parsep 0mm
\item $S \rightarrow B$; rate $k_1(N,t)=q_1(t)(N_B-N)$,
\item $B \rightarrow S$; rate $k_{-1}(N)=q_{-1}N$, 
\item $P \rightarrow B$; rate $k_{-2}(N,t)=q_{-2}(t)(N_B-N)$,
\item $B \rightarrow P$; rate $k_2(N)=q_2N$.
\end{enumerate} 
We allow $q_1$ and $q_{-2}$ to undergo a slow periodic modulation with
a frequency $\omega$, which can be achieved in the biochemical context
by coupling $S$ and $P$ to particle baths with modulated chemical
potentials. In other transport problems, such as transportation
systems, the same modulation may be produced by time-of-day
variations. We note that, unlike in \cite{pilgram-03}, our formulation
has three time scales: fast instantaneous jumps among states,
equilibration in the bin, and adiabatic changes of the rates.

Now the path integral technique of \cite{pilgram-03} can be
applied. Since $N\gg1$, there exists a time scale $\delta t$, over
which many transitions into and out of $B$ happen, but the fractional
change in the bin occupancy remains small, $1\ll\delta N\ll N$. Then
the rate changes $\delta k_i$, $i=-2,-1,1,2$ are also small, and all
transitions are uncorrelated and Poissonian. Thus the probability of the
number of
particle transitions for the $i$'th reaction over time $\delta t$,
denoted by $\delta Q_i$, can be written as $P(\delta Q_{i},t) =
\frac{1}{2\pi} \int_{-\pi}^{\pi} d \chi_{i} e^{-i\chi_{i} \delta Q_{i}
  + N_B H_{i} (\chi_i,t)\delta t}$, where $N_BH_i\delta
t=k_i(N,t)[\exp(i\chi_i)-1]\delta t \equiv k_i(N,t) e_{\chi_i}\delta
t$ is the mgf of a Poisson process with the mean $k_i\delta t$. Note
that we define $e_{x}\equiv e^{ix}-1$ for any $x$.

Our goal is to find the mgf of the net particle number $Q_P$
transfered into $P$ during a long time interval $(0, T)$.  This is
formally given by an integral over fluxes at each moment of
(discretized) time weighted by probabilities $P(\delta Q_i,t)$ and
constrained by particle conservation laws:
\begin{multline}
  \langle e^{i\chi_C Q_{P}}\rangle = \int \prod
  \limits_{k=1}^{T/\delta t} dN(t_k) \prod \limits_{i=\pm 2,\pm1} 
  d\delta Q_i(t_k) P[\delta Q_i(t_k)]\times
  \displaybreak[0]
 \\ \times e^{i\chi_C
    (\delta Q_2(t_k) -\delta Q_{-2}(t_k)) }  \;
  \delta [N(t_{k+1})-N(t_k) \\ - \delta Q_1 (t_k) - \delta Q_{-2}(t_k)+
  \delta Q_{-1} (t_k)+\delta Q_{2}(t_k) ].
\label{path1}
\end{multline} 
Here we used the identity $Q_P= \sum_{k=1}^{T/\delta t}
[\delta Q_2(t_k) -\delta Q_{-2}(t_k)]$, and we introduced a variable
$\chi_C$, which is conjugated to $Q_P$ and ``counts'' particle
transfers into/out of $P$.  Now, using the Fourier representation of
the $\delta$-function, 
we integrate over $\delta Q_i(t_k)$ and $\chi_i(t_k)$ in (\ref{path1})
and reduce it to a path integral over $N$ and its conjugated $\chi$
\begin{equation}
  \langle e^{i\chi_C  Q_{P}}\rangle = 
  \int DN(t)  D\chi(t) 
  e^{\int_{0}^{T} dt \left( i\chi \dot{N} +N_BH( \chi,N,t)\right)},
\label{path2}
\end{equation}
where all pre-factors are absorbed into the measure, and
\begin{multline}
  H(\chi,N,t)=\left[q_1(t)e_{-\chi}+q_{-2}(t)e_{-(\chi+\chi_C)}\right](1-N/N_B)\\+
  \left[q_{-1}e_{\chi}+q_2e_{(\chi+\chi_C)}\right]N/N_B.\label{heffY}
\end{multline}
The explicit time-dependence of
$H$ is due to the slow periodic changes in $q_1(t)$ and $q_{-2}(t)$.

The exponent in (\ref{path2}) has a factor of $N_B$ in it. Thus for
$N_B\to\infty$, the path integral is dominated by the {\em saddle
  point} or {\em classical} values $\chi_{{\rm cl}}$ and $N_{{\rm
    cl}}$,
\begin{equation}
  i\dot{\chi}_{{\rm cl}}=\frac{\partial H(\chi_{{\rm cl}},
    N_{{\rm cl}},t)}{\partial N_{{\rm cl}}},\,\,\,\,\,\,\,
  i\dot{N}_{{\rm cl}}=-\frac{\partial H(\chi_{{\rm cl}},
    N_{{\rm cl}},t)}{\partial \chi_{{\rm cl}}}.  
\label{semicl1}
\end{equation}
Here the boundary terms disappear, as explained in detail in
\cite{pilgram-04}. Furthermore, since the Hamiltonian (\ref{heffY}) is
linear in $N$ there are no higher order in $1/N$ corrections.

Since we assume adiabatic variation of $q_1$ and $q_{-2}$, the
quasi-equilibrium is a good approximation to the exact solution of
(\ref{semicl1}).  It corresponds to setting time-derivatives in
(\ref{semicl1}) to zero and treating time as a parameter.  For
periodic rates variations, this leads to
\begin{eqnarray}
  e^{-i\chi_{{\rm cl}}} &\approx& \frac{K_-+
    \sqrt{K^2+4q_1q_2e_{\chi_C}+4q_{-1}q_{-2}e_{-\chi_C}} }
  {2(q_{1}+q_{-2}e^{-i\chi_C})}, 
\label{semicl2a}
\\
N_{{\rm cl}} &\approx& \frac{N_B(q_1+q_{-2}e^{-i\chi_C})}{q_1+q_{-2}e^{-i\chi_C}
  +(q_{-1}+q_2e^{i\chi_C}) e^{2i\chi_{{\rm
        cl}}}},  
    \label{semicl2b}
\end{eqnarray}
where $\approx$ denotes the accuracy of $O(\omega/q_i)$,
$K=q_1+q_{-2}+q_{-1}+q_2$, $K_-=q_1+q_{-2}-q_{-1}-q_2$.
Since the Hamiltonian is quadratic in its arguments near the saddle
point, corrections of the order $O(\omega/q_i)$ in
(\ref{semicl2a},~\ref{semicl2b}) lead to $O[(\omega/q_i)^2]$
contributions to the mgf, setting the accuracy of our results. We now
have
\begin{equation}
  \langle e^{i\chi_C  Q_{P}}\rangle \approx
  e^{N_B\left[{\int_{{\bf c}} {\bf A} \cdot d {\bf q} +
      \int_{0}^{T} H( \chi_{{\rm cl}},N_{{\rm cl}},t)dt}\right]},
\label{path4}
\end{equation}
where the vector ${\bf A}$, $A_i=i\chi_{{\rm cl}} ( \partial_{q_i}
N_{{\rm cl}}) /N_B$, is defined in the space of parameters $q_i$, and the
contour ${\bf c}$ is given by $q_i(t)$.  For the periodic driving, as
we consider here, with a period $T_0=2\pi/\omega$ and with fixed
$q_{-1}$ and $q_2$, we rewrite the contour integral as the integral of
$F_{q_1,q_{-2}}=\partial_{q_1} A_{{-2}}- \partial_{q_{-2}} A_{1}$ over
the surface ${\bf S_c}$ enclosed by ${\bf c}$. Then
\begin{equation}
Z \equiv \langle e^{i\chi_C  Q_{P}}\rangle \approx e^{N_B S_{\rm
    geom}+N_BS_{\rm cl}},
\label{path5}
\end{equation}
where 
\begin{align}
&S_{{\rm geom}}=\frac{T}{T_0}\oint_{{\bf c}}{\bf A} \cdot d {\bf q} =\frac{T}{T_0} \int_{{\bf S_c}} dq_1 dq_{-2} F_{q_1,q_{-2}}({\bf
  q}),
\label{sgeom}\\
&F_{q_1,q_{-2}}({\bf
  q})=-\frac{e_{-\chi_C}(e^{i\chi_C}q_2+q_{-1})}{[4q_1q_2e_{\chi_C}+
  4q_{-1}q_{-2}e_{-\chi_C}+K^2]^{3/2}},
\label{BC}\\
&S_{\rm cl}=\frac{-T}{2T_0} \int_0^{T_0} dt \left\{ K-
  \sqrt{K^2+4q_1q_2e_{\chi_C}+4q_{-1}q_{-2}e_{-\chi_C}} \right\}.
\label{scl}
\end{align}
The 2-form $F_{q_1,q_{-2}}({\bf q})$ is an analog of the Berry
curvature in quantum mechanics. As follows from (\ref{sgeom}), nonzero
Berry curvature is responsible for the reversible component in the
particle fluxes. Its presence in our model is due to particle
exclusion within the central bin.  If $k_{1}$ and $k_{-2}$ were
independent of $N$, $F_{q_1,q_{-2}}$ would be zero.

Now moments of the flux between absorbing states can be derived easily
by differentiating (\ref{path5}) with respect to $\chi_C$. In
particular, the average flux is
\begin{multline}
  J=J_{\rm pump}+J_{\rm cl}=N_B\left[\iint_{\bf  S_c} dq_1dq_{-2} 
  \frac{q_2+q_{-1}}{T_0K^3}\right.\\+
 \left.\int_0^{T_0} dt\, \frac{q_1q_{2}-q_{-1}q_{-2}}{KT_0}\right],
\label{JJ}
\end{multline}
where the pump term is due to the particle interactions and the
corresponding geometric contribution, while the classical flux would
exist even in the stationary limit.  Notice that this flux is $N_B$
times its value for a single driven Michaelis-Menten enzyme, studied
in \cite{sinitsyn-07}. The same holds for the entire mgf, and hence
for all flux moments. Thus we refer the reader to \cite{sinitsyn-07}
for further analysis of the model. Here we note that this scaling is
not a coincidence since the current model is, indeed, equivalent to
$N_B$ independent enzymes, where $N$ corresponds to the number of 
enzyme-substrate complexes.

In \cite{sinitsyn-07}, we used an analogy with the quantum mechanical
Berry phase to derive the pump flux (\ref{JJ}), while a classical
stochastic treatment, which does not require diagonalizing large
matrices, is being used now.  Existence of these alternative
approaches is not surprising because any discrete quantum mechanical
system can be mapped onto a mathematically equivalent classical
Hamiltonian system \cite{heslot-85}, and then the Berry phase
transforms into a dynamic contribution to the classical action
\cite{liu-03}.  The contribution of the present derivation is to show
that one can derive the classical Hamiltonian for a discrete Markov
chain by considering many identical independent copies of the system
and identifying the per-copy contribution by taking the large copy
number limit.  Alternatively, one can derive the classical
reformulation directly from the Schr\"odinger's equation as well
\cite{heslot-85,liu-03}.

{\em The reversible ratchet effect.}  Now we show that the geometric
contribution to the mgf is responsible also for the ratchet effect in
a periodic potential.  Consider a system of noninteracting particles
moving in a periodic potential $V(x,t)$, which changes adiabatically
with time so that $V(x,t)=V(x,t+T_0)$ and $V(x,t)=V(x+L,t)$.  In the
overdamped case, the average density of particles satisfies the
Fokker-Plank equation
\begin{equation}
\partial_t \rho (x,t) = -\partial_x [A(x,t) \rho(x,t)]
+ D \partial_x^2 \rho (x,t),
\label{FP}
\end{equation}
where $D$ is the diffusion coefficient, and $A(x,t)=-\partial_x
V(x,t)$ is the force.  The current in this model under an adiabatic
deformation of the potential was previously studied in
\cite{parrondo-98}, and the similarity of the final expression and the
Berry phase in quantum mechanics was pointed out.  The close
connection between the classical ratchets and the Berry phase also has
been anticipated in \cite{shi,astumian-02pt}.  In our following
rederivation, we explicitly show that the ratchet current has its
origins in the geometric phase. Namely, it emerges from the complex
geometric phase of the particle flux mgf.

To study diffusion without the external field, $A(x,t)=0$,
Refs.~\cite{pilgram-04} derived the path integral for the
mgf by discretizing the space into small intervals of length $a \ll
L$, indexed by $i$. Then Poisson transition rates among the
neighboring intervals are prescribed in a way that the continuous
limit $a\rightarrow 0$ recovers the diffusion equation. This reduces
the path integral derivation to an already solved problem of
stochastic transitions among a discrete set of states.  To include the
force $A(x,t)$, we assume that it creates an asymmetry in the left and
right transition rates. For example, (\ref{FP}) can be recovered if
the transition rates are such that during a short time $\delta t$ the
average numbers of particles transfered left and right are $\la \delta
Q_{i\to i-1}\ra=D\rho(x_i)\delta t /a$ and $\la \delta Q_{i\to
  i+1}\ra=D\rho(x_i)\delta t /a+A(x_i)\rho(x_i)\delta t$,
respectively.  Then repeating the same steps as in \cite{pilgram-04}
and taking the continuous limit, we find the following path integral
representation of the generating function:
\begin{equation}
  Z=\langle e^{i\chi_C  Q}\rangle = 
  \int D\rho (x,t)   D\chi(x,t) \,
  e^{\int_{0}^{T_0} dt \int_0^L  dx\left[ i\chi \dot{\rho} +H(\rho, \chi)\right]},
\label{pathd}
\end{equation}
where $Q$ is the difference between the numbers of particles passing
through $x=0$ in the right and the left directions during the period
$T_0$. The Hamiltonian is
\begin{equation}
H(\rho, \chi) = -iA(x,t)\rho \frac{\partial \chi}{\partial x}+
iD \frac{\partial \rho}{\partial x}\frac{\partial \chi}{\partial x} -
D \rho \left( \frac{\partial \chi }{\partial x}\right)^2.
\label{HH}
\end{equation}
The dependence on the counting field $\chi_C$ in (\ref{HH}) is hidden
in the boundary conditions on $\chi$ \cite{pilgram-03}, which, for a
periodic system with the spatial period $L$, are
$\rho(L)=\rho(0)$, and $\chi(L)=\chi(0)-\chi_C$.
Now, solving the saddle point equations and substituting the result
back into the action in the path integral, we write the mgf in a
familiar form $Z(\chi_C)=\exp\{S_{\rm geom}(\chi_C)+S_{\rm
  cl}(\chi_C)\}$, where
$S_{\rm geom}(\chi_C)=\int_{0}^{T_0} dt \int_0^L  dx\,\left( i\chi_{\rm cl} 
\dot{\rho}_{\rm cl} \right)$,
and 
$S_{\rm cl}(\chi_C)=\int_{0}^{T_0} dt \int_0^Ldx\, 
H(\rho_{\rm cl}(\chi_C), {\chi}_{\rm cl}(\chi_C))$. 

The analysis simplifies if we are interested only in mean currents,
rather than in their fluctuations. Then we consider $\chi_C\ll1$ and
find the contribution to $\log Z$ that is linear in it. In fact, only
$S_{\rm geom}$ has this contribution in our case.
To determine it, it is sufficient to find $\rho_{\rm cl}(x,t)$ to the
zeroth order and $\chi_{\rm cl}(x,t)$ to the first order in
$\chi_C$. This results in 
$\rho_{\rm cl}(x,t)\approx[Q_0/R_{-}(t)]e^{-V(x,t)/k_BT}$,
$\chi_{\rm cl}(x,t)\approx[-\chi_C/R_{+}(t)] \int_0^x e^{V(x',t)/k_BT}dx'$,
where $R_{\pm}(t)=\int_0^L e^{\pm V(x,t)/k_BT}dx$, and $Q_0=\int_0^L
\rho_{\rm cl}(x,t)|_{\chi_C=0} dx$ is the number of particles per unit
cell.
This leads to
$Z(\chi_C,T_0)=\exp[i\chi_CJT_0 +O\left(\chi_C^2 \right)$]
, where the terms in $O(\chi_C^2)$ can reveal the higher order
cumulants, 
and the average current $J=-(i/T_0)(\partial_{\chi_C} \log Z)_{\chi_C=0}$ is
\begin{equation}
J= \frac{Q_0}{2T_0} \int_0^{T_0}dt \int_0^L dx\, (\partial_t v\, \partial_x u-
\partial_x v\,\partial_t u),
\label{curr}
\end{equation}
where we introduced $u(x,t)=\frac{1}{R_{-}(t)} \int_0^x
e^{-V(x',t)/k_BT}dx'$ and $v(x,t)=\frac{1}{R_+(t)} \int_0^x
e^{V(x',t)/k_BT}dx'$ and used the property $u(L,t)=v(L,t)=1$.  The
integrand in (\ref{curr}) is a pure curl of a vector ${\bf A}$ with
components $A_x=v(x,t)\partial_x u(x,t)$ and $A_t=v(x,t)\partial_t
u(x,t)$ defined in the two dimensional space-time. Thus the current
can be expressed as 
\begin{equation}
J=\frac{Q_0}{2T_0} \oint_{{\bf c}} {\bf A \cdot dr},
\end{equation}
where ${\bf dr}=(dx,dt)$, and ${\bf c}$ is the contour that encloses a
space-time cell with boundaries at $x=0,L$ and $t=0,T_0$.
  
For a uniformly shifting potential $V(x,t)=V(x-tL/T_0)$, $R_{\pm}$ are
time-independent, and the integration in (\ref{curr}) leads
to $J=Q_0/T_0-(Q_0/T_0)L^2/(R_+R_{-})$.  The first term in this
expression is the quantized contribution which is dominating in the
limit of a large potential amplitude.  In \cite{shi}, this
quantization of the classical ratchet current was connected to the
Chern number of the Bloch band related to the potential $V(x)$.
 

{\em Pump current in the SIS epidemiological model.}  In a final
calculation, we show how the stochastic path integral allows
derivation of pump-like effects in novel scenarios; specifically
where, unlike in our first example, the system cannot be factored into
non-interacting identical stochastic subsystems.  We consider the
standard Susceptible-Infected-Susceptible (SIS) mechanism of an
infection outbreak, which is a good model for influenza. State of the
art epidemiological modeling uses deterministic dynamics \cite{nick1},
which tracks only fractions of populations in various states during an
outbreak progression. However, it is understood that stochasticity may
be essential.  Thus here we discuss if stochasticity, and especially
effects due to slow variability of the infectivity and the recovery
rates, can affect disease outbreaks.

Let's denote infected individuals by $I$ and their number by $N$. The
disease spreads due to a permanent infection source and because it can
be transmitted by the infected individuals.  All infected people
eventually recover.
 
Thus the full kinetic scheme is
\begin{enumerate}\itemsep 0mm \partopsep0mm\topsep 0mm
\item $\emptyset \rightarrow I$; rate $k_1$ (permanent infection
  source);
\item $I \rightarrow \emptyset$; rate per infected individual $k_2$  (recovery);
\item $I \rightarrow I+I$; rate per infected individual $k_3$
  (infection spread by contacts).
\end{enumerate}
Here $k_i$ are independent of $N$ because we assume that outbreaks are
small in comparison to the total population size (still
$N\gg1$ is possible).  This requires that $k_2>k_3$, so that, if
stochasticity is unimportant, the deterministic steady state solution
is $N_{\rm st}=k_1/(k_2-k_3)$, and the stationary flux into and out of
the infected state is $J_{\rm st}=k_2N_{\rm
  st}=k_1k_2/(k_2-k_3)$. This model is a birth-death process, and,
with time-independent rates, it has been studied extensively.  Here we
are interested in estimating (possibly substantial) effects of rate
time-dependence.  The Hamiltonian in the path integral for this model
is
\begin{equation}
H(\chi,N,t)=k_1(t)e_{-\chi}+k_2Ne_{(\chi + \chi_C)}+k_3(t)Ne_{-\chi},
\label{hm}
\end{equation}
where $\chi$ is the conjugated variable to $N$ and $\chi_C$ 
counts the flux out of
$I$.   With $N\gg1$, we can use the saddle point analysis, which
is exact since $H$ is linear in $N$.

Now consider a periodic time dependence of the rates $k_i$, which may
be due to the time-of-day or seasonal effects. For simplicity, we
assume that only $k_1$ and $k_3$ vary, and the recovery rate $k_2$
remains constant. As before, the mgf has
both the classical and the geometric contributions,
i.e. $Z=\exp[S_{\rm cl}+S_{\rm geom}]$.  The classical one is the
average of the stationary mgf 
over the period of the rates variation, $T_0$, while the geometric one
is again an integral over the surface ${\bf S_c}$ inside the contour
enclosed by $k_1(t)$ and $k_3(t)$:
\begin{align}
&S_{\rm cl}+S_{\rm geom}=\frac{T}{T_0}\int_0^{T_0}dt H[\chi_{\rm
  cl}(t),N_{\rm cl}(t),t] \nonumber
\\
&\quad\quad\quad+ \frac{T}{T_0} \iint_{{\bf S_c}}dk_1dk_3F_{k_1,k_3}({\bf k}),
\label{FCS6}\\
&F_{k_1,k_3}({\bf k})=\frac{k_2(K_--2k_3e_{\chi_C}-\kappa)}{2k_3^2\kappa^2},
\label{Bcurv2}
\\
&H(\chi_{\rm cl},N_{\rm cl},t)= \frac{k_1(K_--\kappa)}{2k_3},
\label{FCS5}
\end{align}
where now $K_-=k_2- k_3$, and $\kappa = \sqrt{K_-^2
  -4k_2k_3e_{\chi_C}}$. This corresponds to the mean flux
$ J= J_{\rm pump} +J_{\rm cl}$, where $J_{\rm pump}= \frac{1}{T_0}
\iint_{{\bf S_c}} \frac{dk_1dk_3\,k_2}{K_-^3}$
is the pump flux due to the geometric contribution, and the classical
flux is $J_{\rm cl}=1/T_0 \int_0^T J_{\rm st}dt$. Notice, in
particular, that $J_{\rm pump}\propto K_-^{-3}$, and it can become
very large near $K_-=0$. Fluctuations are easy to compute as well by
differentiating (\ref{FCS6}).


{\em Conclusion.}  Based on the stochastic path integral technique, we
built the theory of geometric fluxes in classical stochastic kinetics,
and we proposed a general approach for identification and calculation
of pump-like currents, including the familiar reversible ratchet, as
well as new phenomena.
In the adiabatic limit, the full counting statistics of pump fluxes is
provided by the term that depends on the choice of the contour in the
parameter space, but does not depend on the rate of the motion along
this contour, and thus has a geometric origin. The stationary saddle
point approximation of the path integral is sufficient for
calculations of this geometric contribution in the case of a large
number of particles.
This approach leads to the complete theory of reversible effects in
nonequilibrium statistical physics.  It will open doors to a study of
such poorly understood systems as ratchets with interacting diffusing
particles, or epidemiological models on complex social networks with
time-dependent parameters.



\begin{acknowledgments} 
We thank M. Wall, F. Alexander, N. Hengartner and R. D. Astumian for useful
discussions and critical reading of this text.
  This work was funded in part by DOE under Contract No.\
  DE-AC52-06NA25396.  IN was further supported by NSF Grant No.\
  ECS-0425850.
\end{acknowledgments}


\begin{thebibliography}{99}

\bibitem{astumian-02pt} R.\ D.\ Astumian and P.\ H\"anggi, {\em Phys.\
    Today} {\bf 55}, 33 (2002);
D.\ Astumian,  {\em
AIP Conf. Proc.} {\bf 658} 221, (2003).

\bibitem{parrondo-98} J.~M.~R.~Parrondo, {\em Phys.\ Rev.\ E} {\bf 57},
  7297 (1997).

\bibitem{shi} Y.\ Shi and Q.\ Niu, {\em Europhys.\ Lett.} {\bf 59}, 324
  (2002).



\bibitem{westerhoff-86} H.\ V.\ Westerhoff et al., {\em
    Proc. Natl. Acad. Sci.\ U.S.A.} {\bf 83}, 4734 (1986);
 V.\ S.\ Markin et al.,
 {\em
    J. Chem. Phys.} {\bf 93}, 5062 (1990);
 R.\ D.\ Astumian et al.,  {\em
   Phys. Rev. A} {\bf 39}, 6416 (1988).



\bibitem{sinitsyn-07} N.\ A.\ Sinitsyn and I.\ Nemenman, {\em EPL}
  {\bf 77}, 58001 (2007).

\bibitem{metrology} J.\ L.\ Flowers and B. W. Petley, {\em Rep. Prog. Phys.} {\bf 64},
1191 (2001)

\bibitem{wilczek-88}
A.\ Shapere and F.\ Wilczek, {\em Phys. Rev. Lett.} {\bf 58}, 2051 (1987);
 A.\ Shapere and F.\ Wilczek, {\em J.\ Fluid Mech.}
  {\bf 198}, 557 (1988).

\bibitem{tsong} T.\ Y.\ Tsong and C.\ H.\ Chang, {\em AAAPS Bulletin} {\bf 13}, 12 (2003).

\bibitem{kagan-91} 
M.\ L.\ Kagan, T.\ B.\ Kepler and I.\ R.\ Epstein,
  {\em Nature} {\bf 349}, 506 (1991).

\bibitem{reimann-02} P.\ Reimann, {\em Phys.\ Rep.} {\bf 361}, 57 (2002).;
 F.\ J\"{u}licher, A.\ Ajdari and J.\ Prost,
  {\em Rev.\ Mod.\ Phys.} {\bf 69}, 1269 (1997);
 R.\ D.\ Astumian and I.\ Derenyl, {\em Eur.\
    Biophys.\ J.} {\bf 27}, 474 (1998).




\bibitem{MM} L.\ Michaelis and M.\ L.\ Menten, {\em Biochem. Z.} {\bf
    49}, 333 (1913).

\bibitem{pilgram-03} S.\ Pilgram et al., {\em Phys.\ Rev.\ Lett.} {\bf
    90}, 206801 (2003).

\bibitem{pilgram-04} A.\ N.\ Jordan, E.\ V.\ Sukhorukov and S.\
  Pilgram, {\em J.\ Math.\ Phys.} {\bf 45}, 4386 (2004);
V.\ Elgart and A.\ Kamenev, {\em Phys. Rev. E}
  {\bf 70}, 051205 (2004).
 
\bibitem{heslot-85} A.\ Heslot, {\em Phys.\ Rev.\ D} {\bf 31}, 1341
  (1985);
S.\ Weinberg, {\em Ann.\ Phys.} {\bf 194}, 336
  (1989).

\bibitem{liu-03} J.\ Liu, B.\ Wu and Q.\ Niu, {\em Phys.\ Rev.\ Lett.}
  {\bf 90}, 170404 (2003);
 B.\ Wu, J.\ Liu and Q.\ Niu, {\em Phys.\ Rev.\ Lett.}
  {\bf 94}, 140402 (2005).

\bibitem{nick1} C.\ Castillo-Chavez et al., {\em Mathematical
    Approaches for Emerging and Reemerging Infectious Diseases:
    Introduction to models, methods, and theory} (Springer: Berlin,
  2006).

\end{thebibliography}
\end{document}